# Schizophrenia research under the framework of predictive coding: body, language, and others

Lingyu Li [1,2], Chunbo Li [*,1,2]

[1] Shanghai Mental Health Center, Shanghai Jiao Tong University School of Medicine, Shanghai 200030, China; [2] Shanghai Jiao Tong University School of Medicine, Shanghai 200023, China

[*] Correspondence: licb@smhc.org.cn (Chunbo Li)

## Abstract

Although there have been so many studies on schizophrenia under the framework of predictive coding, works focusing on treatment are very preliminary. A model-oriented, operationalist, and comprehensive understanding of schizophrenia would promote the "therapy turn" of further research. We summarize predictive coding models of embodiment, co-occurrence of over- and under-weighting priors, subjective time processing, language production or comprehension, self-or-other inference, and social interaction. Corresponding impairments and clinical manifestations of schizophrenia are reviewed under these models at the same time. Finally, we discuss why and how to inaugurate a therapy turn of further research under the framework of predictive coding.

**Healing the *schizo-* of schizophrenia research**

Despite that psychiatry progressed surprisingly, schizophrenia, the most challenging mental disorder, remains rising prevalence and causes substantial burden of disease (1). The challenges are partly from an underlying "schizo-" among nowadays research, where neurobiological deficits, objective symptoms, and subjective experience are somewhat disconnected. Over the past decades, the framework of predictive coding has attracted inexhaustible interests and been enriched by stably increasing studies from neuroscience and psychopathology. Predictive coding was originally raised for the sensory processing of visual cortex (2). In the hierarchical model, perception is composed of predictions based on prior beliefs (or *priors*) and incoming sensory signals (Figure 1.a). To minimize prediction errors, beliefs update according to signals mismatching with original predictions (error signals), and cortical neurons are thus regarded as residual error detectors. Detailed mathematical and logical basis of predictive coding was reviewed by Millidge et al (3). Incorporating with theories like active inference, Bayesian inference, free-energy principle, etc. into its framework, predictive coding appears to address aforesaid disconnects — — symptoms (positive symptoms, negative symptoms, and disorganization) arise from inappropriate inferences (Bayesian beliefs updating) induced by neurobiological abnormalities (of e.g., dopamine, synaptic plasticity, and neural oscillations) (4-7). Recently, several researchers propose accounts of psychiatric interventions under the framework of predictive coding (8-10). These preliminary works could herald an exciting "therapy turn" in further research on schizophrenia.

To facilitate the "therapy turn", this review offers a model-oriented, operationalist, and comprehensive understanding of schizophrenia. Every day, we interact with others through bodies and language, but something goes wrong in schizophrenia, impairing patient's abilities. This review, therefore, is structured by these three domains—body, language, and others. We synthesized relevant models, suggested many promising issues, and finally discussed a roadmap for further research.

**Body: shortcut of consciousness**

*Embodiment and disembodiment*

Through body, mind interacts with the world directly, and this process is known as *embodiment,* by which human become embodied subjects. Unlike visual perception, bodily movement involves perception-motor (or *sensorimotor*) process and requires more contextual predictions, error detection, and belief updating (Figure 1. b). Further, execution of a complete task can be conceptualized as a succession of these processes over time (Figure 1. c). Phenomenologically, tight coupling sensorimotor processing is related to sense of selfhood, while a disturbed one can induce psychotic symptoms (11). In schizophrenia, clinicians noticed disembodiment as one core feature (12). From perspective of neuroscience, disembodiment could be mismatch between motor intentions and multisensory feedback, failure of multisensory integration, or failure of body representations (13). Consistently, abnormal movements are identified as the first observable neurological risk factor of schizophrenia (14, 15). And some suspect that motor dysfunction is primary disturbance of schizophrenia other than cognitive deficits (16, 17).

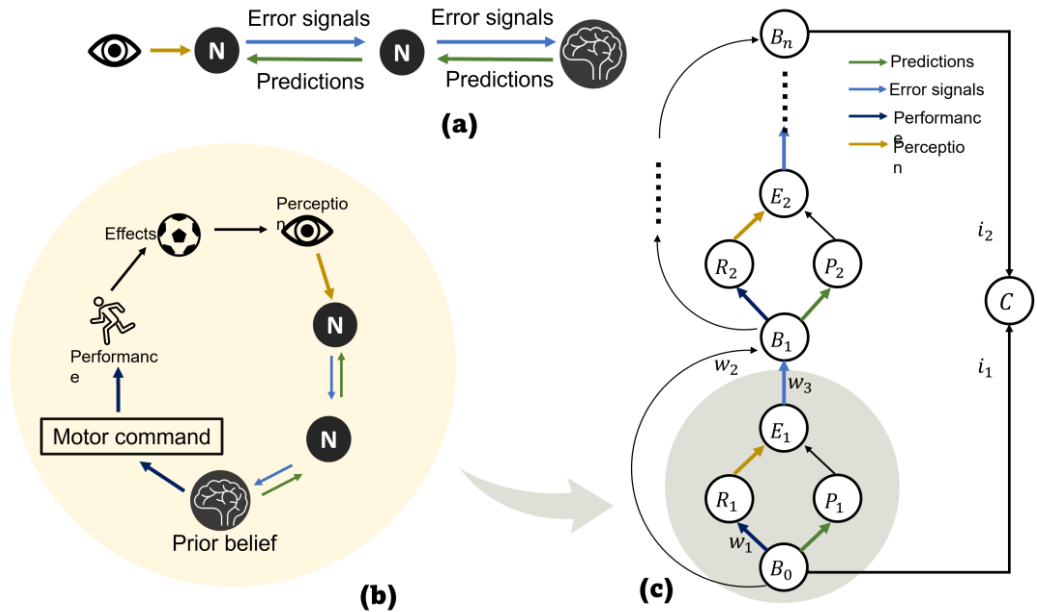

Figure 1. Prediction coding accounts of perception and sensorimotor processing. (a) Hierarchical model of perception, which is composed of predictions and incoming error signals; (b) Model of sensorimotor processing coupling perception and motor. (c) Task execution. $B_j$ provides intentions of action and corresponding predictions $P_{j+1}$; $B_j$ to $R_{j+1}$ denotes implementing of intentions, and $w_1 \in [0,1]$ is proportional to degree of execution; Error signals $E_j$ are propagated back to update beliefs, with precisions $w_2$ and $w_3$; the task finished, consistence ($C$) of results ($B_n$) and original intentions ($B_0$) is assessed retrospectively, acting as a comparator.

However, the longitudinal process from primary motor dysfunction to schizophrenia is difficult to examine in human brain. A plausible explanation, mapped onto figure 1. c, is that primary motor dysfunction corresponds to decreased $w_1$, which would cause increased error signals ($E$) and finally decreased $C$. Since decreased $w_1$ hinders one to improve results, a "backpropagation" of other parameters would probably occur, and consequently, belief updating is more dependent on priors than error signals. Interestingly, Annabi et al proposed a neural architecture involving two recurrent neural networks (motor and visual) in line with predictive coding, and investigated the bidirectional interactions between motor and visual modules (18). This illuminating work provides a paradigm to simulate above longitudinal process for computational neuroscience.

Consequence of primary motor dysfunctions includes over-weighting priors and under-weighting error signals. However, real weights of these two items in schizophrenia are more complicated and even paradoxical when considering specific symptoms.

*How do over- and under-weighting priors co-occur?*

Due to its generation from interaction of predictions and sensory inputs, perception is viewed as "controlled hallucination" (19). While aforesaid overweighting priors could cause genuine hallucinations because one tends to perceive according to priors rather than sensory inputs (20-24). But another view argues that hallucinations are due to weaker priors and abnormal amplification of neural signals (e.g., amplified inner speech causes auditory hallucination) (4). Similarly, empirical research found that both stronger and weaker priors correlate with delusion proneness (25-29). Patients with schizophrenia may reveal stronger and weaker priors at the same time, but how is this possible?

Researchers resolve the contradiction by multilevel hierarchical predictive coding models, which are roughly divided into low- and high-level systems, bearing the function of perception and inference, respectively. False perceptions by disrupted low-levels require compensations (or explanations) from inferences of high-levels, so over- and under-weighting priors coexist in different hierarchies (30). Sterzer et al further suggest a looser hierarchy where low- and high-level systems are controlled partly independently, which accounts for higher heterogeneity of psychosis (4). Different from single hierarchical perspective, Leptourgos et al argue that above two systems are parallel (authors named them ego- and allo-centric system), receiving inputs from same sensory systems (31). Activities of dopamine and related neural circuits may provide neurobiological bases, including abnormal temporo-limbic and prefrontal-striatal connectivity, and co-occurrence of reduced phasic dopamine response and increased spontaneous phasic dopamine release (6, 32).

To summarize, the paradox is dealt in spatial dimension, that is, over- or under-weighting priors are located in their respective positions. But challenges remain, for example, increased and decreased sense of agency could co-occur (i.e., they overlap spatially). A temporal dimension may help to separate these spatial overlaps.

*Toward a logical time*

Subjective time processing could be seen as a collider of [past → present ← future], a combination of prospective expectation and retrospective judgment (33-35). Conditional on actions in a certain instant, Riemer distinguished two directions: forward in time to generate predictions (effects) and backward in time to form post-hoc inferences of intentions (causes) (36). When the forward model goes wrong, retrospective elaborations are responsible for causality inferences (e.g., "I would not have intended to do that!") (37). Absence of intention inference and enhanced binding of action and effects account for co-occurrence of two paradoxical experiences: (1) there's no intention responsible for my movements, so I feel like being manipulated; (2) my movements indeed cause things happening, so these effects should be attributed to my movements (36, 38, 39).

Mapping to figure 1.c, for one with motor dysfunctions, compensation of decreased $C$ derives a "backpropagation" of $i_1$ and $i_2$, which correspond to impaired [past → present] inference and enhanced [present ← future] attribution, respectively. Interestingly, these findings are quite in keeping with the Lacanian psychoanalytic term *logical time*, which refutes lineal time and emphasizes that retroaction and anticipation work in tandem to influence the being of human (40).

**Language: long road of interpretation**

*"There is no direct apprehension of the self ⋯⋯through the short cut of consciousness but only by the long road of interpretation of signs." (41)*

Manifesting flawed thought, language disorders are core symptoms of schizophrenia. As natural language processing techniques burgeoning, language disorders have been a promising marker for schizophrenia (42, 43). Neuroscientific linguistics, meanwhile, yielded encouraging findings utilizing noninvasive tools. Nevertheless, predictive coding accounts of language remain an arduous task due to multiple hierarchies of language (phonetics, phonology, morphology, syntax, semantics, and pragmatics) and complex characteristics of thoughts (such as generality and rich compositionality), which draws a critical concern about whether predictive coding is a feasible model for language and thought (44-46).

*Predictive coding accounts of language*

Despite six hierarchies of language as mentioned, neurobiological focus can be placed on three parts: phonetics, semantics, and pragmatics (Figure 2). Phonetically, language production is effects of articulatory muscles and thus shares brain regions with motor controls (47-49). Speaking is thus also a sensorimotor process depicted in previous section. Brain regions including cerebellum, Broca's area, and Wernicke's area provide articulatory motor command and generate predictions of auditory perceptions, and then error signals are fed back (49-51). Interestingly, some researchers think of these predictions partly serving as *inner speech* (52, 53). And because of affinities between motor and speech, *embodied language* in both schizophrenia and healthy people attracts increasing research interests (54-56).

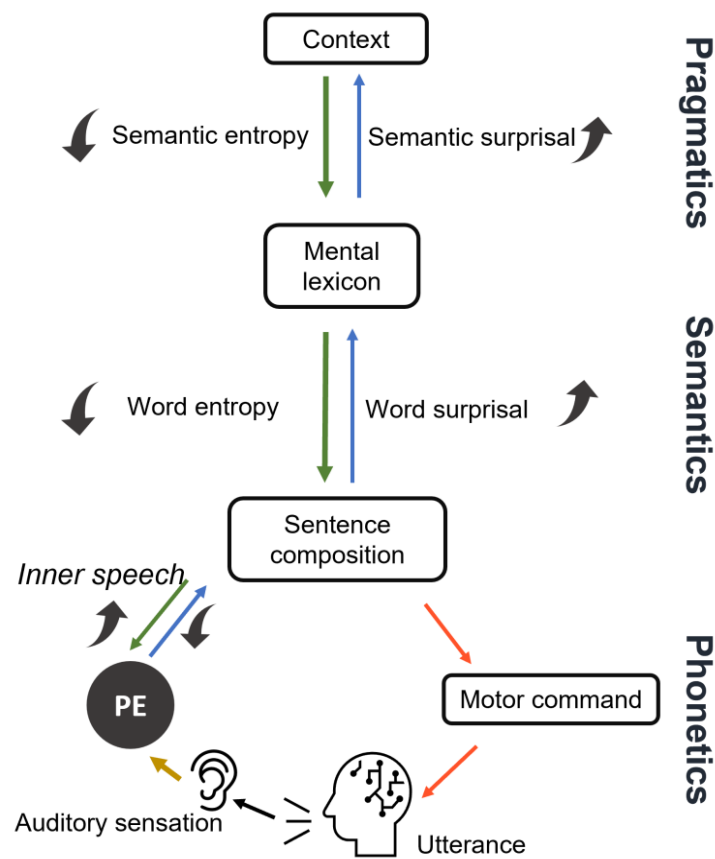

Figure 2. Predictive coding accounts of language, involving three hierarchies: phonetics, semantics, and pragmatics.

Semantically, considering sentence as a sequence of words appearing sequentially, predictions and error signals can be quantified by *word entropy* and *word surprisal*, respectively. Based on long-term semantic knowledge (or *mental lexicon*), *word entropy*, quantifying probability distribution of upcoming words, measures uncertainty of the next word. *Word surprisal*, mathematically defined as the negative logarithm of conditional probability, denotes how newsworthy an observed word is (i.e., *amount of information*) (46, 57). Imaging studies distinguished different brain regions sensitive to word surprisal and entropy among healthy subjects during natural language comprehension and explained them in framework of predictive coding (58-60). Neural oscillations tracking findings suggest that β- and γ-band power reflects semantical predictions and error signals, respectively, and altered β-γ coupling might result in aberrant semantics in schizophrenia (7, 61). Sharpe et al found that reduced N400 amplitudes were related to lower word entropy (62). Altogether, during sentence generation or comprehension, the probabilistic distribution of upcoming words is predicted according to semantic relatedness between previous words and all words in mental lexicon (computational model of *semantic priming*), and error signals are propagated back (46).

Pragmatics studies meaning generation among speaker and hearer in a dynamic context, and focus on linguistic phenomena including deixis, presupposition, implicature, and speech acts (63). *Context* is a global grasp of what, when, where, and who are involved in current linguistic condition, and hence requires lots of cognition control (64). S*emantic entropy* and *semantic surprisal* quantify predictions and error signals at pragmatic level (65, 66). Semantic entropy constrains mental lexicon in current context, known as *semantic inhibition*——otherwise, computations of word entropy would be too huge to give predictions (46, 67, 68). Error signals (semantic surprisal) collect unexpected information to update context through cognition control (69, 70). As for updating of context based on prior and posterior belief, *relative entropy* could be a competent tool (71). Overall, language production is tightly coupled with motor, perception, memory, and cognition, so disorders of language would easily turn one's world chaotic.

*Sound and fury: language disorders in schizophrenia*

Many studies suggested that aberrant cognitive language processing contributes to auditory verbal hallucinations (AVH) even more than disturbed auditory perceptions (72-74). Under predictive coding framework, Wilkinson distinguished three subtypes: inner speech AVH, memory-based AVH, and hypervigilance AVH (75). Given articulation as a sensorimotor process, AVH could be attributed to stronger priors at phonetics level (20, 22, 76-78). Unstable memory is another main concept in cognitive mechanisms of AVH, which endorses that hallucinators mis-integrate intrusive memories into current context, and AVH is thus conceptualized as failure of accessing context cues (79-81). Impaired cognitive inhibition (or inhibitory control) may contribute to the noisy memory-based contexts because irrelevant memories are activated involuntarily (82). Despite altered top-down processes providing AVH materials, they cannot explain the external biases of AVH, which implicates deficits of self-monitoring or source-monitoring (83). Disrupted bottom-up processing may arrest discrimination between self- and other-generated voices (84, 85). Rather than intentional "misattribution", Heinks-Maldonado et al argued that "misperception", uncertainty about the source of voices, may depict AVH more accurately (86). In summary, AVH does possess phenomenological diversity and correlates to disorders in both up-down and bottom-up processing at multiple levels.

Formal thought disorders (FTD), consisting of positive *disorganization* and negative *impoverishment*, anchors in impairments at semantic and pragmatic levels such as semantic hyper-/hypo-priming, and absence of contextual guidance (67). Little information, higher preservation, smaller clusters, and higher switching rates characterize hyper-priming in schizophrenia (87, 88). Hyper-priming could be explained through impaired cognition and subsequent failure of context access (89). From perspectives of predictive coding, disturbed pragmatic predictions (i.e. increased semantic entropy) represent broaden semantic distribution, which is called *semantic disinhibition*, an accepted etiology of hyper-priming (89). To compensate for difficulties in semantic retrieval or arrangement due to impaired predictions, weights of semantic surprisal are upregulated, which consequently leads to higher semantic similarity, just as Alonso-Sánchez et al suggested (90).

Besides cognitive deficits, there are also abnormalities in automatic semantic processing. Impaired N400 modulation and reduced β-band oscillations reflect underweighting semantic predictions (i.e., increased word entropy) (61, 89). Among numerous studies denoting ongoing debates, we do appreciate works of Sharpe et al, where semantic priming was examined utilizing two prime-target blocks with varied predictive validity (10% and 50%) (62). For the lower predictive validity block, semantic priming was minimal in both patients with schizophrenia and healthy subjects. But when transited to higher predictive validity block without telling participants, patients showed hypo-priming and impaired N400 modulation contrasted to control group. The authors proposed that patients with schizophrenia represented a "rigid" Bayesian brain with under-weighting predictions. In another experiment, they examined lexical alignment in schizophrenia and put forward that preserved (or sometimes enhanced) bottom-up process balanced out top-down deficits in semantic processing (91). Altogether, FTD could result from disorders at semantic and pragmatic levels. Phenomenologically, patients are plunged into a linguistic world full of uncertainties, and no wonder that some found correlations between language disorders and bad emotions like anhedonia and hostility (92, 93).

*Delusions as subtle FTD?*

It is doubtful but attractive to consider linguistic contributions to delusion. Hinzen et al distinguished two subsets of delusions: *referential delusions* where personal significances are attributed to neutral events, and *propositional delusions* where patients make false assertions without attribution (94). Then they proposed that although literally understandable (e.g., *I am Jesus*), propositional delusions represent failure of grammatical meaning where patients wrongly embed arguments under relevant predicates (e.g., *being Jesus*). Furthermore, their fMRI study reported impaired deixis processing in schizophrenia with and without AVH, which possibly contributes to referential delusions (94, 95). The idea regarding delusions as subtle FTD accounts for why delusions are bizarre and cannot be corrected by solid evidence. While it remains understudied if delusions correlate with other linguistic phenomena like presupposition, implicature, and speech act. And this somewhat postmodern idea merits further attention and research.

**Other: live in intersubjectivity**

One cannot be aware of self without recognizing others, so schizophrenia, always regarded as disorders of self, is also disorders of other. Under the framework of predictive coding, in this section are discussed two parts: how brain differentiates self and other, and how human interact with other based on theory of mind (ToM).

*Self or other?*

Many schizophrenic symptoms like AVH, thought insertion, control delusion, Cotard delusion, etc. signify failures of self-or-other inference. Rubber hand illusion (RHI) suggests that even healthy people could misrecognize external objects as "mine", and predictive coding tells what happened (96). More importantly, RHI indicates there is no solid self but competition of self-or-other in brain (e.g., actual and rubber hand), and predictive coding could serve as the comparator distinguishing them. Classical comparator model is embedded in sensorimotor processing. In more detail, intention & prediction, prediction & real-time result, and intention & final result are compared during movement, contributing to sense of initiating, predictions updating, and sense of agency, respectively (53). After comparison, possibility distributions of motor performer are generated (96). Among healthy people, sense of agency could be reduced through top-down interference, i.e., underweighting predictions (97, 98).

In the case that our brain involves multiple hierarchies of predictive coding, self-or-other inference would implement hierarchically too. This multi-level structure could account for mixed findings in RHI of patients with schizophrenia (Figure 3). Most RHI studies reported significant group differences between schizophrenia and healthy subjects in subjective rating scales, but not in objective proprioceptive drift, and RHI in schizophrenia cannot be explained by perceptual alterations (99). Instead, severity of delusions and aberrant spatial body perception (impaired high-level priors) may increase sensitivity to RHI (99, 100). In schizophrenia patients, weights of higher-level priors tend to decrease, which is in line with their high sensitivity to RHI. Sense of agency is more complicated in sensorimotor and language production, and the "logical time" might play important role as discussed above.

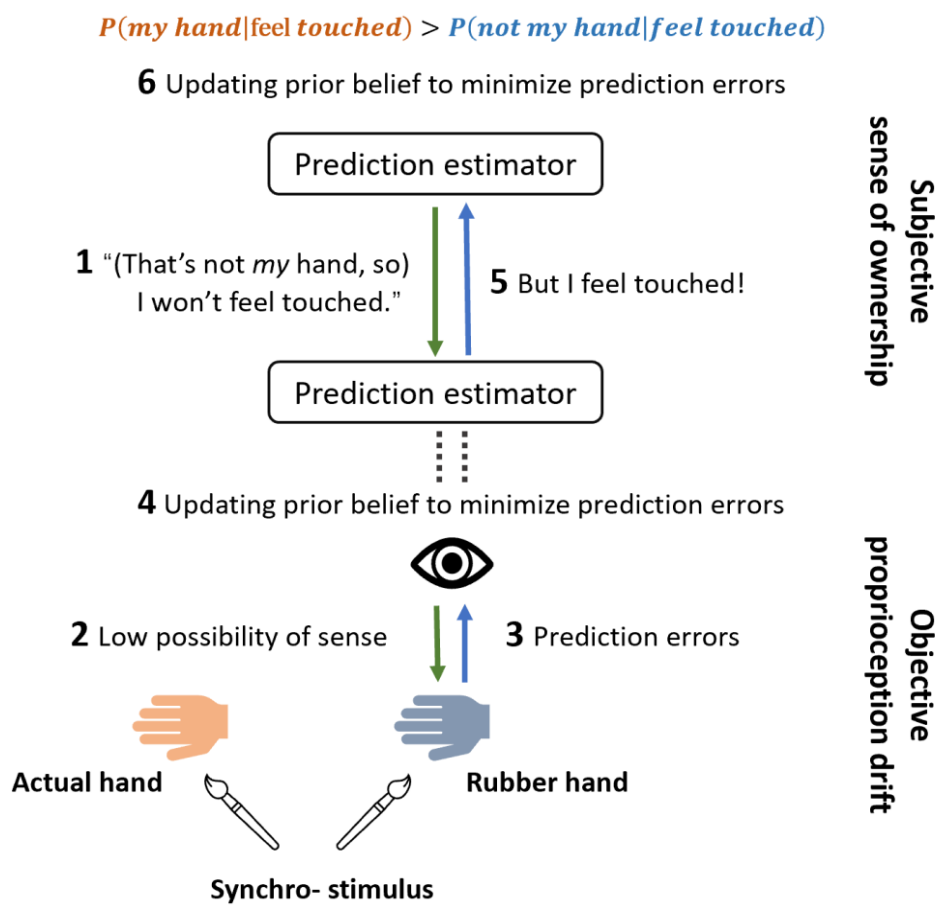

Figure 3. Predictive coding accounts of rubber hand illusion.

Anyhow, as Parnas et al suggested, although many complain of long-lasting disorders of self, complete forms of human existence are preserved (101). It seems that brain squashes a "self" or "other" into unoccupied "body", or in other words, we construct ourselves through everyday life, which would be an interesting question.

*Theory of mind and social interaction*

ToM refers to the ability by which one realizes that others are also individuals who act, speak, think, love, hate, etc., and understands their situations, that is, ability to live in intersubjectivity, phenomenologically. The neurobiological basis of ToM is mirror neurons, and predictive coding has been extended to modeling ToM (102, 103). On timescales of seconds, minutes, and long-term, three types of predictions are generated including others' observable movements, beliefs & desires, and preferences & personalities, for which different brain regions are responsible (104). These ToM predictions might be implemented hierarchically in order of stable personalities, concrete intentions, and finally specific movements (105, 106). Bottom-up feedbacks arise from exteroceptive, proprioceptive, and interoceptive senses (5, 107). Empirical research examined the predictive coding model by analyzing fMRI responses of ToM brain regions in movie-viewing or story-reading tasks (108-110). However, most studies focused on healthy subjects, and differences between schizophrenia and healthy people require further research.

Mirza et al formulated a computational model to simulate emotion recognition according to social cues and suggested two approaches to performance of schizophrenia including imprecise top-down guidance and abnormal bottom-up cues access (111). Consistently, Stuke et al reported the correlations between psychosis proneness and enhanced priors for detecting socially meaningfully visual stimuli (112). Diaconescu et al introduced a computational model of intention inference, suggesting that imbalanced precisions of priors and error signals contribute to generation and consolidation of persecutory delusions (113). Jeganathan et al proposed a multiple hierarchical model combining social inferences with personal behaviors and studied negative symptoms under this model (5). They held that imprecise higher-level social predictions result in uncertainty about social milieu and thus a series of phenomena, including affective blunting, demotivating social actions, and blurry self-other boundaries.

**Concluding remarks and future perspectives**

This review summarizes theoretical and empirical studies under the framework of predictive coding, provides an operationalist understanding of schizophrenia, and suggests some promising issues. Researchers are excited to witness that predictive coding is closing chasms between phenomenological psychopathology and neuroscientific theories, biological and computational psychiatry, and neuroscience and artificial intelligence. However, critics argue that this theory is nearly not falsifiable, and being addicted to this framework discourages the development of other equally effective alternatives (45, 114). From a perspective of Lakatos's philosophy of science, we believe that predictive coding is becoming a *research programme*, of which the incontestable hardcore is an actively predicting brain. To keep (also examine) it a progressive research programme, we must do more than interpreting manifestations of schizophrenia. Therefore, the effects of a therapy turn of further research under predictive coding are two-folds: (1) potential clinical benefit of people with schizophrenia, and (2) reinvigoration of this framework through testable inferences.

Predictive coding prepares a robust and integrative framework where almost any intervention can get theoretical support. A promising roadmap could be comprehensive computational models of human-being (treating body, language, and other as a triad), accounts of how interventions like antipsychotics, psychotherapy, and behavioral training (motor, language, and theory of mind) improve one's predictive coding, and then mutual computational models involving different treatments. Finally, an ideal therapy may combine a predictive coding model of patients, understandings of their subjective experiences, and corresponding treatments on abnormal predictive coding process. The most inspiring suggestion might be to create a "digital twin" of patient computationally, which allows speculating the effects of intervention (115). All in all, this framework is far from well-developed, still calling for further interdisciplinary collaboration.

**Acknowledgement**

None.

**Author contributions**

Lingyu Li: Conceptualization, methodology, writing, and visualization.

Chunbo Li: Conceptualization, methodology, and project administration.

**Competing interest**

No competing interests were reported.

**References**

1. Charlson FJ, Ferrari AJ, Santomauro DF, Diminic S, Stockings E, Scott JG, et al. Global Epidemiology and Burden of Schizophrenia: Findings From the Global Burden of Disease Study 2016. Schizophrenia bulletin. 2018;44(6):1195-203.
2. Rao RP, Ballard DHJNn. Predictive coding in the visual cortex: a functional interpretation of some extra-classical receptive-field effects. 1999;2(1):79-87.
3. Millidge B, Seth A, Buckley C. Predictive Coding: a Theoretical and Experimental Review2021.
4. Sterzer P, Adams RA, Fletcher P, Frith C, Lawrie SM, Muckli L, et al. The Predictive Coding Account of Psychosis. Biological psychiatry. 2018;84(9):634-43.
5. Jeganathan J, Breakspear M. An active inference perspective on the negative symptoms of schizophrenia. The Lancet Psychiatry. 2021;8.
6. Maia TV, Frank MJ. An Integrative Perspective on the Role of Dopamine in Schizophrenia.

Biological psychiatry. 2017;81(1):52-66.
7. van Pelt S, Heil L, Kwisthout J, Ondobaka S, van Rooij I, Bekkering H. Beta- and gamma-band activity reflect predictive coding in the processing of causal events. Social cognitive and affective neuroscience. 2016;11(6):973-80.
8. Lutz A, Mattout J, Pagnoni G. The epistemic and pragmatic value of non-action: a predictive coding perspective on meditation. Current Opinion in Psychology. 2019;28:166-71.
9. Adams RA, Vincent P, Benrimoh D, Friston KJ, Parr T. Everything is connected: Inference and attractors in delusions. Schizophrenia research. 2022;245:5-22.
10. Putica A, Felmingham KL, Garrido MI, O'Donnell ML, Van Dam NT. A predictive coding account of value-based learning in PTSD: Implications for precision treatments. Neuroscience & Biobehavioral Reviews. 2022;138:104704.
11. Poletti M, Tortorella A, Raballo A. Impaired Corollary Discharge in Psychosis and At-Risk States: Integrating Neurodevelopmental, Phenomenological, and Clinical Perspectives. Biological psychiatry Cognitive neuroscience and neuroimaging. 2019;4(9):832-41.
12. Fuchs T, Schlimme JE. Embodiment and psychopathology: a phenomenological perspective. Current opinion in psychiatry. 2009;22(6):570-5.
13. Giummarra MJ, Gibson SJ, Georgiou-Karistianis N, Bradshaw JL. Mechanisms underlying embodiment, disembodiment and loss of embodiment. Neuroscience & Biobehavioral Reviews. 2008;32(1):143-60.
14. van Harten PN, Walther S, Kent JS, Sponheim SR, Mittal VA. The clinical and prognostic value of motor abnormalities in psychosis, and the importance of instrumental assessment. Neuroscience & Biobehavioral Reviews. 2017;80:476-87.
15. Filatova S, Koivumaa-Honkanen H, Hirvonen N, Freeman A, Ivandic I, Hurtig T, et al. Early motor developmental milestones and schizophrenia: A systematic review and meta-analysis. Schizophrenia research. 2017;188:13-20.
16. Izawa J, Asai T, Imamizu H. Computational motor control as a window to understanding schizophrenia. Neuroscience research. 2016;104:44-51.
17. Abboud R, Noronha C, Diwadkar VA. Motor system dysfunction in the schizophrenia diathesis: Neural systems to neurotransmitters. European Psychiatry. 2017;44:125-33.
18. Annabi L, Pitti A, Quoy M. Bidirectional interaction between visual and motor generative models using Predictive Coding and Active Inference. Neural Networks. 2021;143:638-56.
19. Paolucci C. Perception as Controlled Hallucination. In: Paolucci C, editor. Cognitive Semiotics: Integrating Signs, Minds, Meaning and Cognition. Cham: Springer International Publishing; 2021. p. 127-57.
20. Horga G, Schatz KC, Abi-Dargham A, Peterson BS. Deficits in Predictive Coding Underlie Hallucinations in Schizophrenia. The Journal of Neuroscience. 2014;34(24):8072.
21. Powers AR, Mathys C, Corlett PR. Pavlovian conditioning–induced hallucinations result from overweighting of perceptual priors. 2017;357(6351):596-600.
22. Corlett PR, Horga G, Fletcher PC, Alderson-Day B, Schmack K, Powers AR, 3rd. Hallucinations and Strong Priors. Trends in cognitive sciences. 2019;23(2):114-27.
23. Gawęda Ł, Moritz S. The role of expectancies and emotional load in false auditory perceptions among patients with schizophrenia spectrum disorders. European archives of psychiatry and clinical neuroscience. 2021;271(4):713-22.
24. Salomon R, Progin P, Griffa A, Rognini G, Do KQ, Conus P, et al. Sensorimotor Induction of

Auditory Misattribution in Early Psychosis. Schizophrenia bulletin. 2020;46(4):947-54.
25. Stuke H, Weilnhammer VA, Sterzer P, Schmack K. Delusion Proneness is Linked to a Reduced Usage of Prior Beliefs in Perceptual Decisions. Schizophrenia bulletin. 2019;45(1):80-6.
26. Baker SC, Konova AB, Daw ND, Horga G. A distinct inferential mechanism for delusions in schizophrenia. Brain : a journal of neurology. 2019;142(6):1797-812.
27. Diaconescu AO, Wellstein KV, Kasper L, Mathys C, Stephan KE. Hierarchical Bayesian models of social inference for probing persecutory delusional ideation. Journal of abnormal psychology. 2020;129(6):556-69.
28. Hauke DJ, Roth V, Karvelis P, Adams RA, Moritz S, Borgwardt S, et al. Increased Belief Instability in Psychotic Disorders Predicts Treatment Response to Metacognitive Training. Schizophrenia bulletin. 2022;48(4):826-38.
29. Sheffield JM, Suthaharan P, Leptourgos P, Corlett PR. Belief Updating and Paranoia in Individuals With Schizophrenia. Biological Psychiatry: Cognitive Neuroscience and Neuroimaging. 2022.
30. Fletcher PC, Frith CD. Perceiving is believing: a Bayesian approach to explaining the positive symptoms of schizophrenia. Nature Reviews Neuroscience. 2009;10(1):48-58.
31. Leptourgos P, Corlett PR. Embodied Predictions, Agency, and Psychosis. Frontiers in big data. 2020;3:27.
32. Heinz A, Murray GK, Schlagenhauf F, Sterzer P, Grace AA, Waltz JA. Towards a Unifying Cognitive, Neurophysiological, and Computational Neuroscience Account of Schizophrenia. Schizophrenia bulletin. 2019;45(5):1092-100.
33. Kent L, Nelson B, Northoff G. Can disorders of subjective time inform the differential diagnosis of psychiatric disorders? A transdiagnostic taxonomy of time. Early intervention in psychiatry. 2022;n/a(n/a).
34. Matthews WJ, Meck WH. Temporal cognition: Connecting subjective time to perception, attention, and memory. Psychological bulletin. 2016;142:865-907.
35. Hogendoorn H. Perception in real-time: predicting the present, reconstructing the past. Trends in cognitive sciences. 2022;26(2):128-41.
36. Riemer M. Delusions of control in schizophrenia: Resistant to the mind's best trick? Schizophrenia research. 2018;197:98-103.
37. Martin J-R. Experiences of activity and causality in schizophrenia: When predictive deficits lead to a retrospective over-binding. Consciousness and cognition. 2013;22(4):1361-74.
38. Voss M, Moore J, Hauser M, Gallinat J, Heinz A, Haggard P. Altered awareness of action in schizophrenia: a specific deficit in predicting action consequences. Brain : a journal of neurology. 2010;133(10):3104-12.
39. Hauser M, Knoblich G, Repp BH, Lautenschlager M, Gallinat J, Heinz A, et al. Altered sense of agency in schizophrenia and the putative psychotic prodrome. Psychiatry research. 2011;186(2):170-6.
40. Evans D. An introductory dictionary of Lacanian psychoanalysis: Routledge; 2006.
41. Ricoeur P. The conflict of interpretations: Essays in hermeneutics: Northwestern University Press; 1974.
42. Voppel AE, de Boer JN, Brederoo SG, Schnack HG, Sommer I. Quantified language connectedness in schizophrenia-spectrum disorders. Psychiatry research. 2021;304:114130.
43. Tan EJ, Meyer D, Neill E, Rossell SL. Investigating the diagnostic utility of speech patterns in

schizophrenia and their symptom associations. Schizophrenia research. 2021;238:91-8.
44. Hartopo D, Kalalo RT. Language disorder as a marker for schizophrenia. Asia-Pacific psychiatry : official journal of the Pacific Rim College of Psychiatrists. 2021:e12485.
45. Williams D. Predictive coding and thought. Synthese. 2020;197(4):1749-75.
46. Armeni K, Willems RM, Frank SL. Probabilistic language models in cognitive neuroscience: Promises and pitfalls. Neuroscience & Biobehavioral Reviews. 2017;83:579-88.
47. Hertrich I, Dietrich S, Ackermann H. The role of the supplementary motor area for speech and language processing. Neuroscience and biobehavioral reviews. 2016;68:602-10.
48. Ashida R, Cerminara NL, Edwards RJ, Apps R, Brooks JCW. Sensorimotor, language, and working memory representation within the human cerebellum. Human brain mapping. 2019;40(16):4732-47.
49. Moberget T, Ivry RB. Prediction, Psychosis, and the Cerebellum. Biological psychiatry Cognitive neuroscience and neuroimaging. 2019;4(9):820-31.
50. Price CJ, Crinion JT, Macsweeney M. A Generative Model of Speech Production in Broca's and Wernicke's Areas. Frontiers in psychology. 2011;2:237.
51. Ford JM. Studying auditory verbal hallucinations using the RDoC framework. Psychophysiology. 2016;53(3):298-304.
52. Jones SR, Fernyhough C. Thought as action: Inner speech, self-monitoring, and auditory verbal hallucinations. Consciousness and cognition. 2007;16(2):391-9.
53. Swiney L, Sousa P. A new comparator account of auditory verbal hallucinations: how motor prediction can plausibly contribute to the sense of agency for inner speech. Frontiers in human neuroscience. 2014;8:675.
54. Zakowicz P, Skibińska M, Pawlak J. Disembodied Language in Early-Onset Schizophrenia. 2022;13.
55. Cayol Z, Rotival C, Paulignan Y, Nazir TA. "Embodied" language processing: Mental motor imagery aptitude predicts word-definition skill for high but not for low imageable words in adolescents. Brain and cognition. 2020;145:105628.
56. Tonna M, Lucarini V, Borrelli DF, Parmigiani S, Marchesi C. Disembodiment and Language in Schizophrenia: An Integrated Psychopathological and Evolutionary Perspective. Schizophrenia bulletin. 2022.
57. Hale J. Information-theoretical Complexity Metrics. Language and linguistics compass. 2016;10(9):397-412.
58. Willems RM, Frank SL, Nijhof AD, Hagoort P, van den Bosch A. Prediction During Natural Language Comprehension. Cerebral Cortex. 2016;26(6):2506-16.
59. Shain C, Blank IA, van Schijndel M, Schuler W, Fedorenko E. fMRI reveals language-specific predictive coding during naturalistic sentence comprehension. Neuropsychologia. 2020;138:107307.
60. Wang L, Schoot L, Brothers T, Alexander E, Warnke L, Minjae K, et al. Predictive coding across the left fronto-temporal hierarchy during language comprehension. Cerebral Cortex. 2022.
61. Meyer L, Lakatos P, He Y. Language Dysfunction in Schizophrenia: Assessing Neural Tracking to Characterize the Underlying Disorder(s)? Frontiers in neuroscience. 2021;15:640502.
62. Sharpe V, Weber K, Kuperberg GR. Impairments in Probabilistic Prediction and Bayesian Learning Can Explain Reduced Neural Semantic Priming in Schizophrenia. Schizophrenia bulletin. 2020;46(6):1558-66.

63. Cap P. Pragmatics, Micropragmatics, Macropragmatics. 2010;6(2):195-228.
64. Chiew K, Braver T. Context Processing and Cognitive Control. 2017. p. 143-66.
65. Macuch Silva V, Franke M. Pragmatic Prediction in the Processing of Referring Expressions Containing Scalar Quantifiers. Frontiers in psychology. 2021;12:662050.
66. Venhuizen NJ, Crocker MW, Brouwer H. Semantic Entropy in Language Comprehension. Entropy [Internet]. 2019; 21(12).
67. Palaniyappan L. Dissecting the neurobiology of linguistic disorganisation and impoverishment in schizophrenia. Seminars in cell & developmental biology. 2021.
68. Augurzky P, Franke M, Ulrich R. Gricean Expectations in Online Sentence Comprehension: An ERP Study on the Processing of Scalar Inferences. Cognitive Science. 2019;43(8):e12776.
69. Brown M, Kuperberg G. A Hierarchical Generative Framework of Language Processing: Linking Language Perception, Interpretation, and Production Abnormalities in Schizophrenia. Frontiers in human neuroscience. 2015;9:643.
70. Haase V, Spychalska M, Werning M. Investigating the Comprehension of Negated Sentences Employing World Knowledge: An Event-Related Potential Study. Frontiers in psychology. 2019;10:2184.
71. Caticha A. Entropy, Information, and the Updating of Probabilities. Entropy (Basel, Switzerland). 2021;23(7).
72. Fuentes-Claramonte P, Soler-Vidal J, Salgado-Pineda P, García-León M, Ramiro N, Santo-Angles A, et al. Auditory hallucinations activate language and verbal short-term memory, but not auditory, brain regions. Scientific reports. 2021;11(1):18890.
73. Conde T, Gonçalves OF, Pinheiro AP. A Cognitive Neuroscience View of Voice-Processing Abnormalities in Schizophrenia: A Window into Auditory Verbal Hallucinations? Harvard review of psychiatry. 2016;24(2):148-63.
74. Matsumoto Y, Ayani N, Kitabayashi Y, Narumoto J. Longitudinal Course of Illness in Congenitally Deaf Patient with Auditory Verbal Hallucination. Case reports in psychiatry. 2022;2022:7426850.
75. Wilkinson S. Accounting for the phenomenology and varieties of auditory verbal hallucination within a predictive processing framework. Consciousness and cognition. 2014;30:142-55.
76. de Boer JN, Linszen MMJ, de Vries J, Schutte MJL, Begemann MJH, Heringa SM, et al. Auditory hallucinations, top-down processing and language perception: a general population study. Psychological medicine. 2019;49(16):2772-80.
77. Alderson-Day B, Moffatt J, Lima CF, Krishnan S, Fernyhough C, Scott SK, et al. Susceptibility to auditory hallucinations is associated with spontaneous but not directed modulation of top-down expectations for speech. Neuroscience of consciousness. 2022;2022(1):niac002.
78. Kafadar E, Fisher VL, Quagan B, Hammer A, Jaeger H, Mourgues C, et al. Conditioned Hallucinations and Prior Overweighting Are State-Sensitive Markers of Hallucination Susceptibility. Biological psychiatry. 2022;92(10):772-80.
79. Ćurčić-Blake B, Ford JM, Hubl D, Orlov ND, Sommer IE, Waters F, et al. Interaction of language, auditory and memory brain networks in auditory verbal hallucinations. Progress in neurobiology. 2017;148:1-20.
80. Badcock JC, Hugdahl K. Cognitive mechanisms of auditory verbal hallucinations in psychotic and non-psychotic groups. Neuroscience & Biobehavioral Reviews. 2012;36(1):431-8.
81. Brébion G, Stephan-Otto C, Cuevas-Esteban J, Usall J, Ochoa S. Impaired memory for

temporal context in schizophrenia patients with hallucinations and thought disorganisation. Schizophrenia research. 2020;220:225-31.

82. Barzykowski K, Radel R, Niedźwieńska A, Kvavilashvili L. Why are we not flooded by involuntary thoughts about the past and future? Testing the cognitive inhibition dependency hypothesis. Psychological research. 2019;83(4):666-83.

83. Brookwell ML, Bentall RP, Varese F. Externalizing biases and hallucinations in source-monitoring, self-monitoring and signal detection studies: a meta-analytic review. Psychological medicine. 2013;43(12):2465-75.

84. Sun Q, Fang Y, Peng X, Shi Y, Chen J, Wang L, et al. Hyper-Activated Brain Resting-State Network and Mismatch Negativity Deficit in Schizophrenia With Auditory Verbal Hallucination Revealed by an Event-Related Potential Evidence. Frontiers in psychiatry. 2020;11:765.

85. Ilankovic LM, Allen PP, Engel R, Kambeitz J, Riedel M, Müller N, et al. Attentional modulation of external speech attribution in patients with hallucinations and delusions. Neuropsychologia. 2011;49(5):805-12.

86. Heinks-Maldonado TH, Mathalon DH, Houde JF, Gray M, Faustman WO, Ford JM. Relationship of Imprecise Corollary Discharge in Schizophrenia to Auditory Hallucinations. Archives of general psychiatry. 2007;64(3):286-96.

87. Kircher T, Stein F, Nagels A. Differences in single positive formal thought disorder symptoms between closely matched acute patients with schizophrenia and mania. European archives of psychiatry and clinical neuroscience. 2022;272(3):395-401.

88. Gabrić P, Vandek M. Semantic fluency reveals reduced functional connectivity between subcategorical co-hyponyms in recent-onset inpatients with first-episode psychosis. Clinical linguistics & phonetics. 2021:1-17.

89. Almeida VN, Radanovic M. Semantic priming and neurobiology in schizophrenia: A theoretical review. Neuropsychologia. 2021;163:108058.

90. Alonso-Sánchez MF, Ford SD, MacKinley M, Silva A, Limongi R, Palaniyappan L. Progressive changes in descriptive discourse in First Episode Schizophrenia: a longitudinal computational semantics study. Schizophrenia. 2022;8(1):36.

91. Sharpe V, Schoot L, Lewandowski KE, Öngür D, Türközer HB, Hasoğlu T, et al. We both say tomato: Intact lexical alignment in schizophrenia and bipolar disorder. Schizophrenia research. 2022;243:138-46.

92. Yu L, Ni H, Wu Z, Fang X, Chen Y, Wang D, et al. Association of Cognitive Impairment With Anhedonia in Patients With Schizophrenia. Frontiers in psychiatry. 2021;12:762216.

93. Dar S, Liebenthal E, Pan H, Smith T, Savitz A, Landa Y, et al. Abnormal semantic processing of threat words associated with excitement and hostility symptoms in schizophrenia. Schizophrenia research. 2021;228:394-402.

94. Hinzen W, Rosselló J, McKenna P. Can delusions be understood linguistically? Cognitive neuropsychiatry. 2016;21(4):281-99.

95. Fuentes-Claramonte P, Soler-Vidal J, Salgado-Pineda P, Ramiro N, Garcia-Leon MA, Cano R, et al. Processing of linguistic deixis in people with schizophrenia, with and without auditory verbal hallucinations. NeuroImage Clinical. 2022;34:103007.

96. Apps MA, Tsakiris M. The free-energy self: a predictive coding account of self-recognition. Neuroscience and biobehavioral reviews. 2014;41:85-97.

97. Martin J-R, Pacherie E. Alterations of agency in hypnosis: A new predictive coding model.

Psychological review. 2019;126:133-52.
98. Andersen M. Predictive coding in agency detection. Religion, Brain & Behavior. 2019;9(1):65-84.
99. Zopf R, Boulton K, Langdon R, Rich AN. Perception of visual-tactile asynchrony, bodily perceptual aberrations, and bodily illusions in schizophrenia. Schizophrenia research. 2021;228:534-40.
100. Prikken M, van der Weiden A, Baalbergen H, Hillegers MH, Kahn RS, Aarts H, et al. Multisensory integration underlying body-ownership experiences in schizophrenia and offspring of patients: a study using the rubber hand illusion paradigm. Journal of psychiatry & neuroscience : JPN. 2019;44(3):177-84.
101. Parnas J, Henriksen MG. Disordered self in the schizophrenia spectrum: a clinical and research perspective. Harvard review of psychiatry. 2014;22(5):251-65.
102. Kilner JM, Friston KJ, Frith CD. Predictive coding: an account of the mirror neuron system. Cognitive Processing. 2007;8(3):159-66.
103. Nour MM, Barrera A. Schizophrenia, Subjectivity, and Mindreading. Schizophrenia bulletin. 2015;41(6):1214-9.
104. Koster-Hale J, Saxe R. Theory of Mind: A Neural Prediction Problem. Neuron. 2013;79(5):836-48.
105. Westra E. Stereotypes, theory of mind, and the action–prediction hierarchy. Synthese. 2019;196(7):2821-46.
106. Tamir DI, Thornton MA. Modeling the Predictive Social Mind. Trends in cognitive sciences. 2018;22(3):201-12.
107. Ondobaka S, Kilner J, Friston K. The role of interoceptive inference in theory of mind. Brain and cognition. 2017;112:64-8.
108. Richardson H, Saxe R. Development of predictive responses in theory of mind brain regions. Developmental Science. 2020;23(1):e12863.
109. Heil L, Colizoli O, Hartstra E, Kwisthout J, van Pelt S, van Rooij I, et al. Processing of Prediction Errors in Mentalizing Areas. Journal of Cognitive Neuroscience. 2019;31(6):900-12.
110. Dungan JA, Stepanovic M, Young L. Theory of mind for processing unexpected events across contexts. Social cognitive and affective neuroscience. 2016;11(8):1183-92.
111. Mirza MB, Cullen M, Parr T, Shergill S, Moran RJ. Contextual perception under active inference. Scientific reports. 2021;11(1):16223.
112. Stuke H, Kress E, Weilnhammer VA, Sterzer P, Schmack K. Overly Strong Priors for Socially Meaningful Visual Signals Are Linked to Psychosis Proneness in Healthy Individuals. Frontiers in psychology. 2021;12:583637.
113. Diaconescu AO, Hauke DJ, Borgwardt S. Models of persecutory delusions: a mechanistic insight into the early stages of psychosis. Molecular psychiatry. 2019;24(9):1258-67.
114. Colombo M, Elkin L, Hartmann S. Being Realist about Bayes, and the Predictive Processing Theory of Mind. 2021;72(1):185-220.
115. Friston K. Computational psychiatry: from synapses to sentience. Molecular psychiatry. 2022.